\documentclass[
    ,final            
  ]
  {aipproc}

\layoutstyle{8x11single}

\usepackage{graphicx}

\begin{document}

\title{Multi-physics Extension of OpenFMO Framework}



\author{Toshiya~Takami}{
  address={Research Institute for Information Technology, Kyushu University,
Fukuoka 812--8581, Japan}
}

\author{Jun~Maki}{
  address={Research Institute for Information Technology, Kyushu University,
Fukuoka 812--8581, Japan}
}

\author{Jun'ichi~Ooba}{
  address={Research Institute for Information Technology, Kyushu University,
Fukuoka 812--8581, Japan}
}

\author{Yuuichi~Inadomi}{
  address={PSI Project Lab., Kyushu University,
3--8--33--710 Momochihama, Sawara-ku, Fukuoka 814--0001, Japan}
}

\author{Hiroaki~Honda}{
  address={PSI Project Lab., Kyushu University,
3--8--33--710 Momochihama, Sawara-ku, Fukuoka 814--0001, Japan}
}

\author{Ryutaro~Susukita}{
  address={Fukuoka IST Foundation, Acros Fukuoka Nishi Office 9F,
1--1--1 Tenjin, Chuo-ku, Fukuoka 810--0001, Japan}
}

\author{Koji~Inoue}{
  address={Department of Informatics, Kyushu University,
6--10--1 Hakozaki, Higashi-ku, Fukuoka 812--8581, Japan}
  ,altaddress={PSI Project Lab., Kyushu University,
3--8--33--710 Momochihama, Sawara-ku, Fukuoka 814--0001, Japan}
}

\author{Taizo~Kobayashi}{
  address={Research Institute for Information Technology, Kyushu University,
Fukuoka 812--8581, Japan}
}

\author{Rie~Nogita}{
  address={Research Institute for Information Technology, Kyushu University,
Fukuoka 812--8581, Japan}
}

\author{Mutsumi~Aoyagi}{
  address={Research Institute for Information Technology, Kyushu University,
Fukuoka 812--8581, Japan}
  ,altaddress={PSI Project Lab., Kyushu University,
3--8--33--710 Momochihama, Sawara-ku, Fukuoka 814--0001, Japan}
}

\begin{abstract}
OpenFMO framework, an open-source software (OSS) platform
for Fragment Molecular Orbital (FMO) method,
is extended to multi-physics simulations (MPS).
After reviewing the several FMO implementations
on distributed computer environments,
the subsequent development planning corresponding to MPS is presented.
It is discussed which should be selected as a scientific software,
lightweight and reconfigurable form or large and self-contained form.
\end{abstract}

\maketitle


\section{Introduction and Overview}

Multi-physics simulations are widely used even in complex scientific studies.
Such calculations are often constructed
by combining multiple theories including
different degrees of approximations and different scales of description.
Since reality and accuracy are required increasingly,
these simulations have become larger and more complicated year by year.
Grids, distributed computer resources over wide-area networks, are expected
to execute such complicated scientific applications,
and have been installed all over the world in order to demonstrate
large-scale heterogeneous simulations with the help of middlewares
\cite{NAREGI05,NAREGI-Web}.
On the other hand, the next generation supercomputer with a peta-scale
performance is already planned in several countries\cite{Riken,PSI-Web}.
Thus, the development of high-performance computing environments
is fast and transient.  As a scientist, it is important to watch the trend of
those computer resources.

In the present contribution,
the multi-physics calculations by Fragment Molecular Orbital (FMO) method
\cite{FMO99}
are constructed on the distributed computing environments.
OpenFMO framework toward ``peta-scale'' computing\cite{OpenFMO,HPCNano06}
is extended to the multi-physics simulations.
It is also discussed what architecture and development policy
should be chosen in the fast-moving world of computing.

\section{Grid-enabled Calculations of FMO}

Before entering the main subject,
we briefly review the grid-enabled FMO implementations
developed in the NAREGI project\cite{NAREGI05}.
These are based
on the famous MO package, GAMESS\cite{GAMESS}.

\subsection{Implementation of a Loosely-coupled FMO}

\begin{figure}
\centering{\includegraphics[scale=0.4]{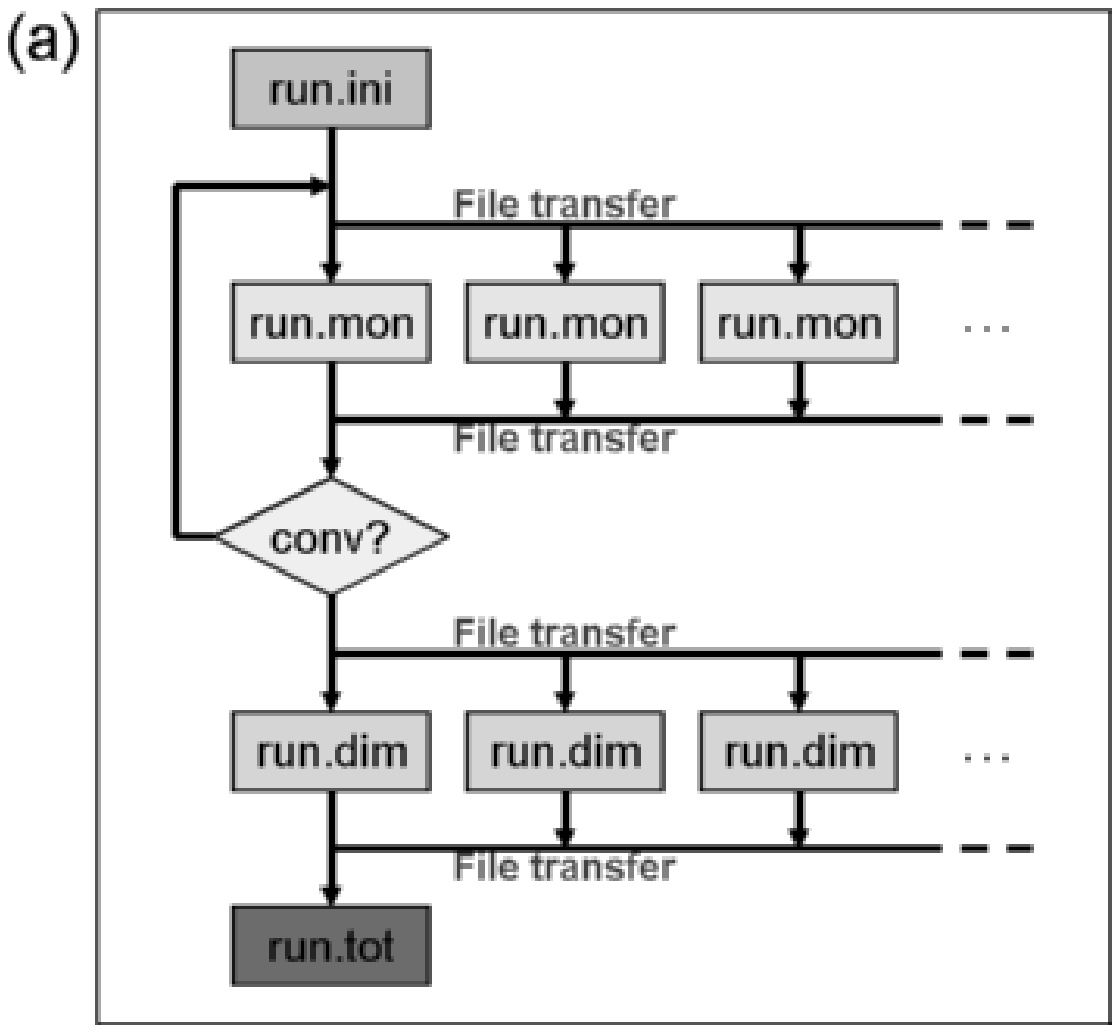}
\includegraphics[scale=0.4]{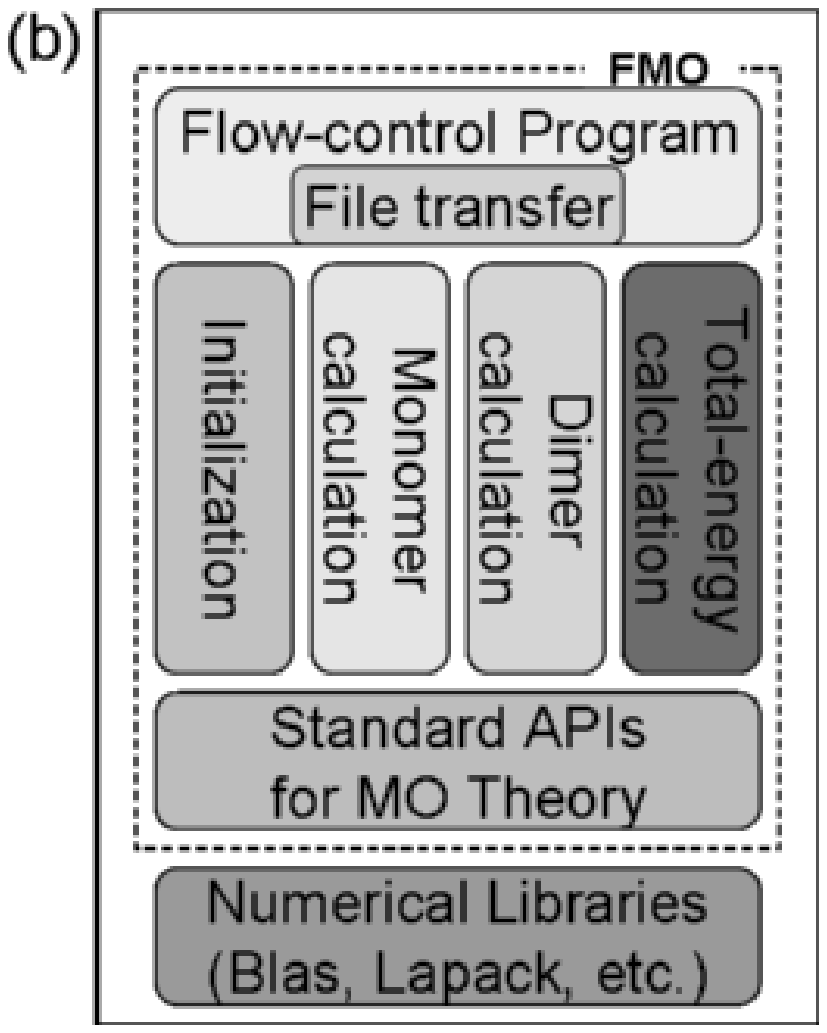}
\includegraphics[scale=0.4]{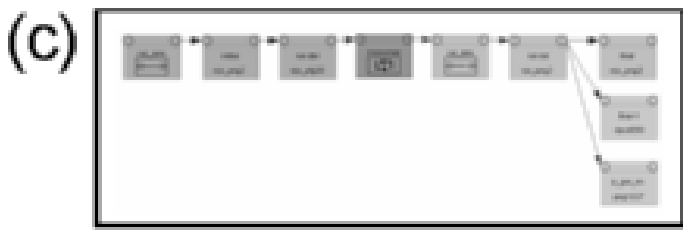}}
\caption{The structure of Loosely-coupled FMO represented in
(a) a flow chart, (b) a program stack,
and (c) graphical icons on the NAREGI Workflow tool.}
\label{fig:lcFMO}
\end{figure}

Although it is usually considered as an approximation to
{\it ab initio\/} molecular orbital (MO) calculations,
the FMO algorithm is a multi-layered problem (see Fig. \ref{fig:lcFMO}(b))
including the MO calculations for each fragment and
the electro-static (ES) interaction between fragments.
In the MO-layer, the quantum mechanical interactions
of all the atoms and electrons within a fragment
are included to obtain a fragment energy.
On the other hand, only the classical ES interaction
is considered when we go over the boundary of fragments.
Since the MO-layer calculations can be executed independently,
we can break the program into loosely-coupled components
corresponding to a large-scale parallel execution
in the distributed computing environments (Fig. \ref{fig:lcFMO}(c)).

\begin{figure}
\centering{\includegraphics[scale=0.33]{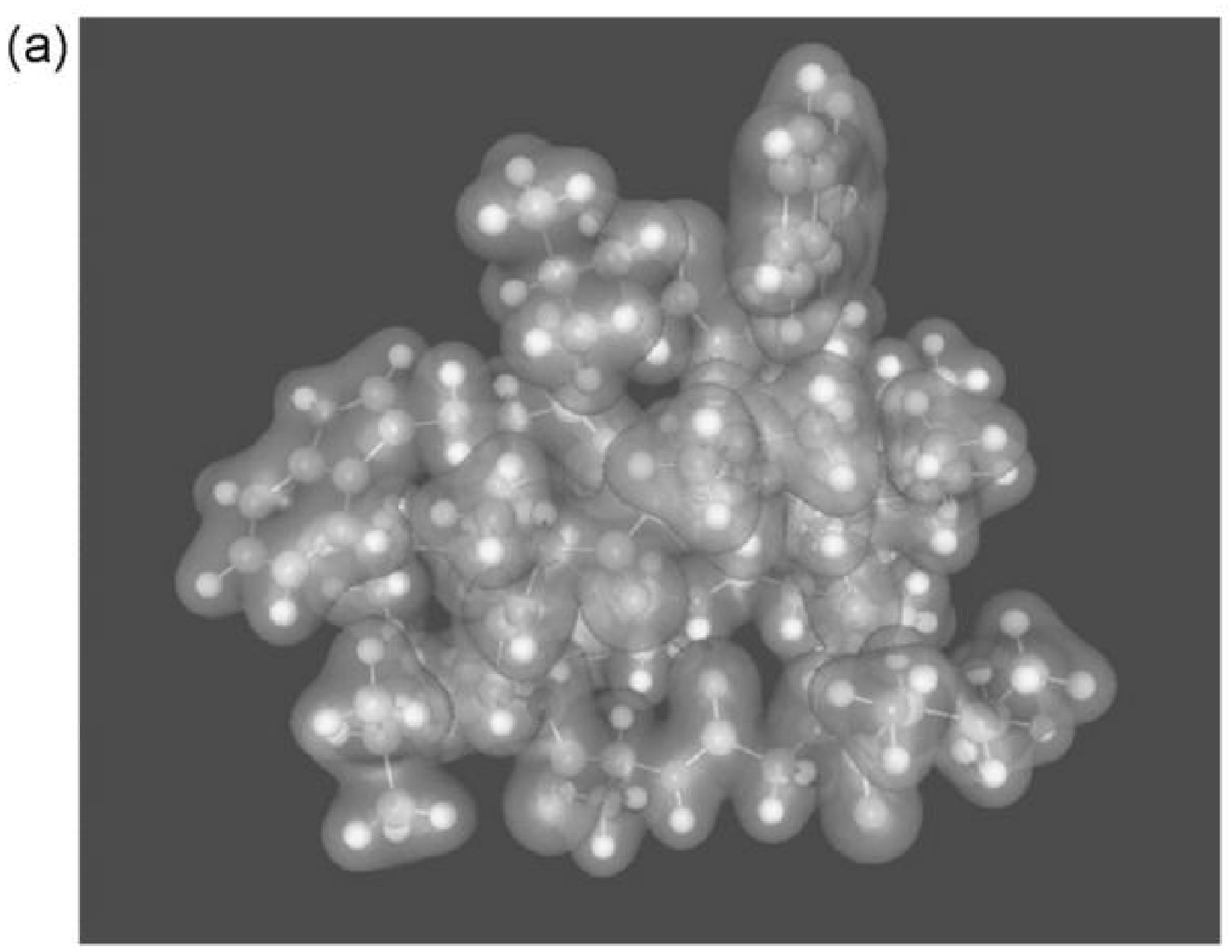}
\includegraphics[scale=0.32]{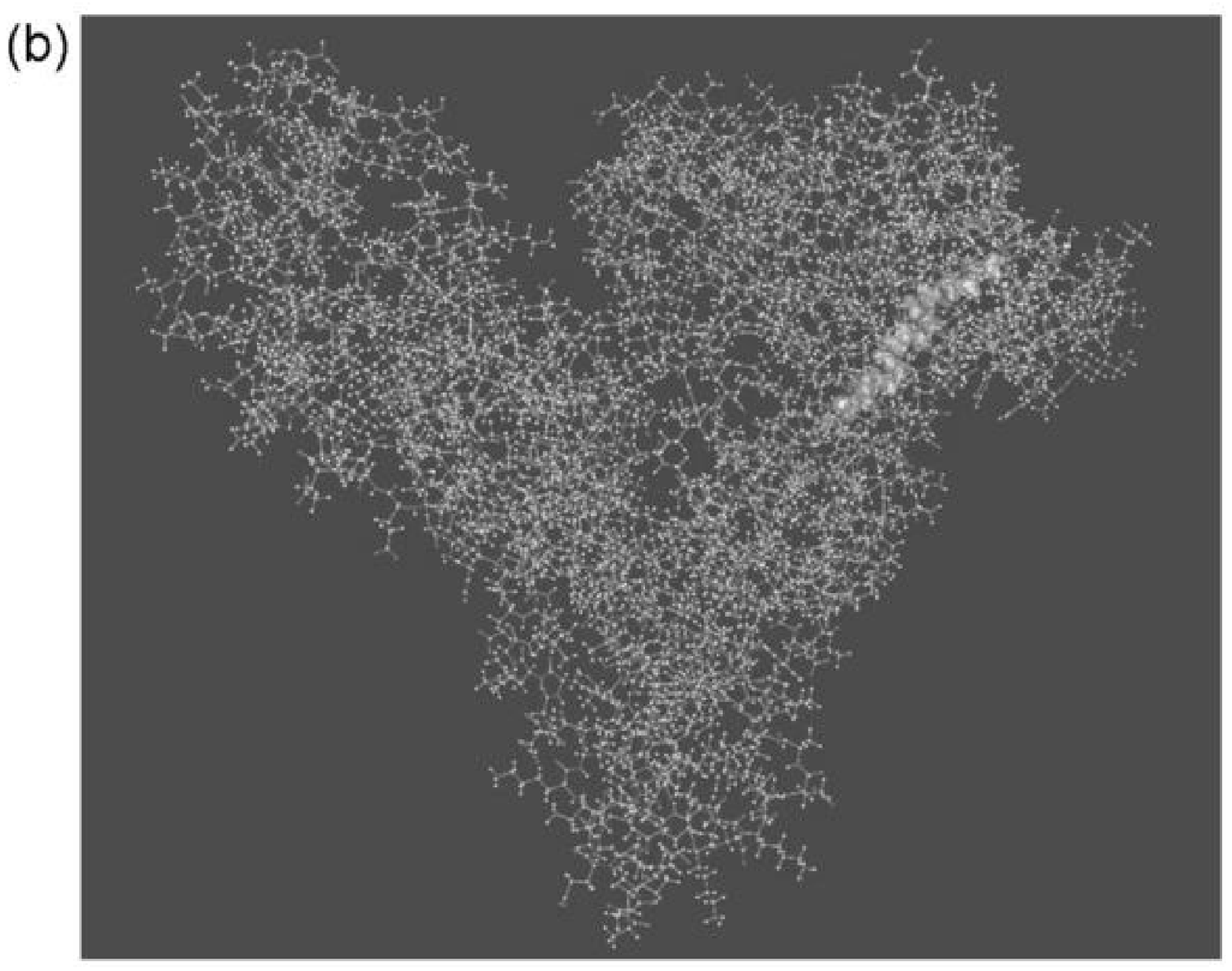}}
\caption{The electron density obtained by FMO
is shown by the use of the NAREGI visualization system:
(a) the total electron density of a Gramicidin-A (1JNO);
(b) the electron density of a fragment in a fatty acid-albumin.}
\label{fig:lcFMO-result}
\end{figure}

The grid-enabled version called ``Loosely-coupled FMO''
was developed as a part of NAREGI\cite{NAREGI05}.
The total control flow is constructed
by the use of the NAREGI Workflow tool.
In Fig. \ref{fig:lcFMO-result} (a),
the total electron density of the whole molecule\cite{Inadomi}
of a Gramicidin-A is shown as an equi-density surface,
and the electron density for one of the fragments in a fatty-acid albumin
is shown in Fig. \ref{fig:lcFMO-result}(b).

\subsection{3D-RISM/FMO Simulation Connected by a Mediator}

As an example of the multi-physics simulations,
a coupled simulation of FMO and 3D-RISM is presented,
where FMO calculations are coupled to
statistical mechanics calculations for molecular liquids
by Reference Interaction Site Model (RISM)\cite{rismBook}.
In order to obtain properties of bio-molecules, drugs, enzymes, etc.,
it is necessary to perform calculations under the influence of a solvent
since these molecules usually work in aqueous solution.
However, the full description of the solute and solvent system
is difficult in general because of the large number of degrees
of freedom.
The standard strategy to solve the problem is to combine,
in some way, originally different theories or programs,
which is the multi-physics approach.

\begin{figure}
\centering{\includegraphics[scale=0.4]{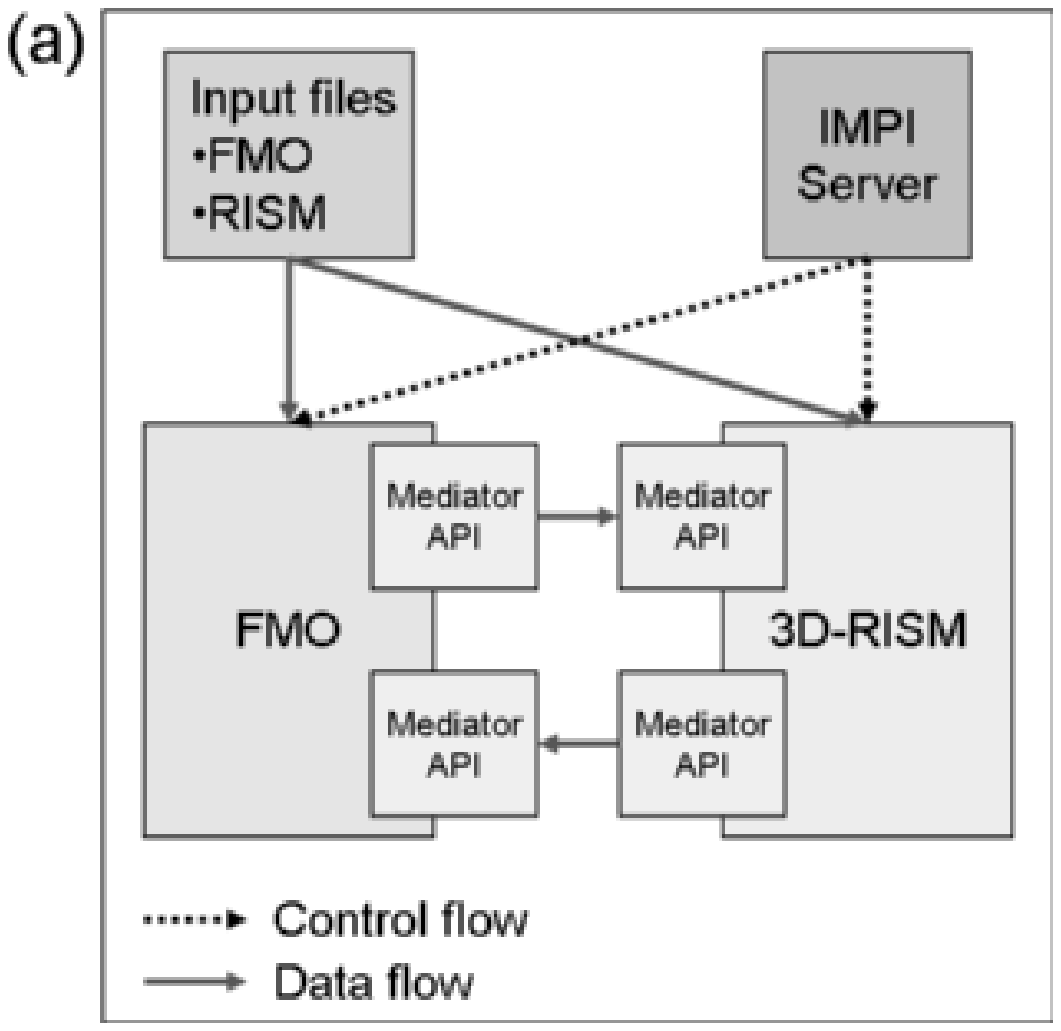}
\includegraphics[scale=0.4]{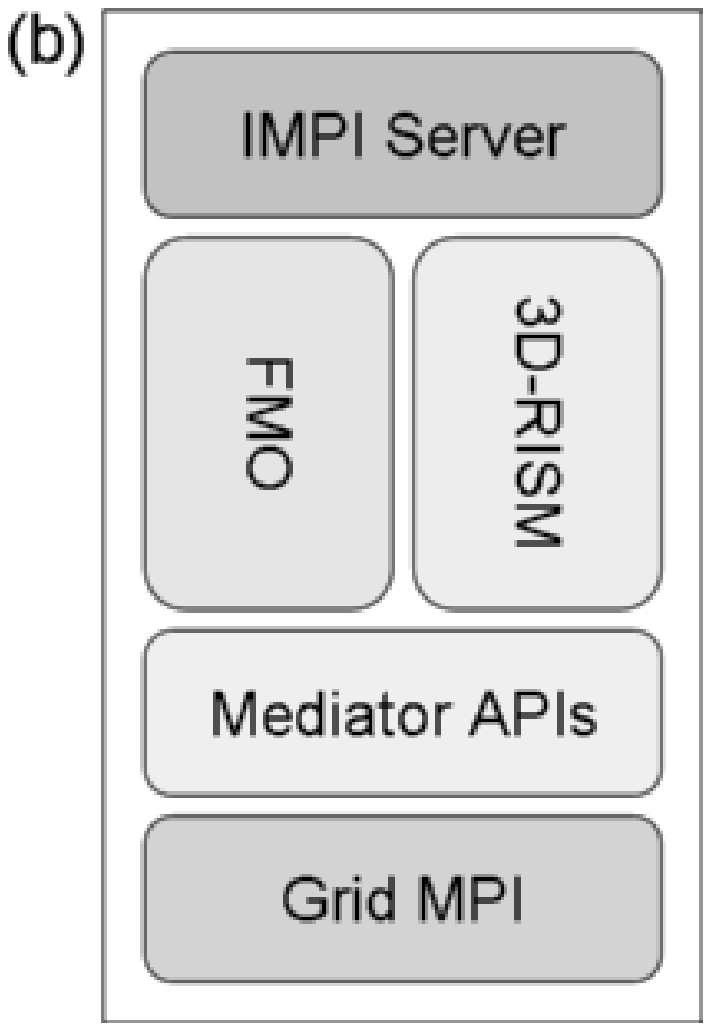}
\includegraphics[scale=0.4]{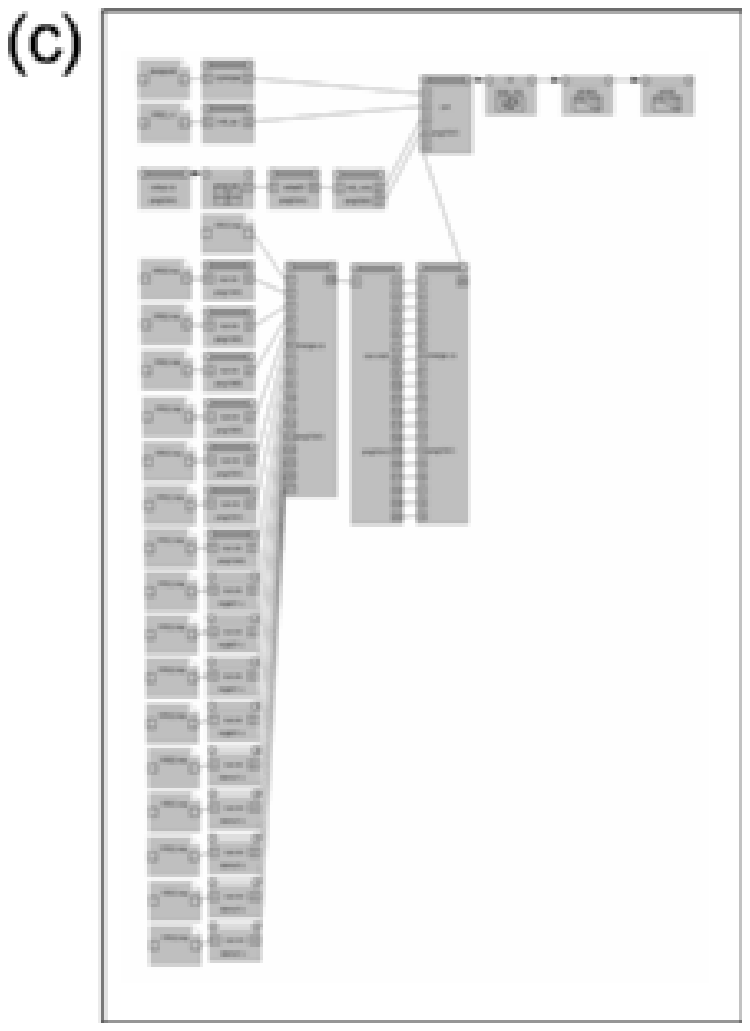}}
\caption{The structure of 3D-RISM/FMO represented in
(a) a flow chart, (b) a program stack,
and (c) graphical icons on the NAREGI Workflow tool.}
\label{fig:rism-fmo}
\end{figure}

\begin{figure}
\centering{
\includegraphics[scale=0.25]{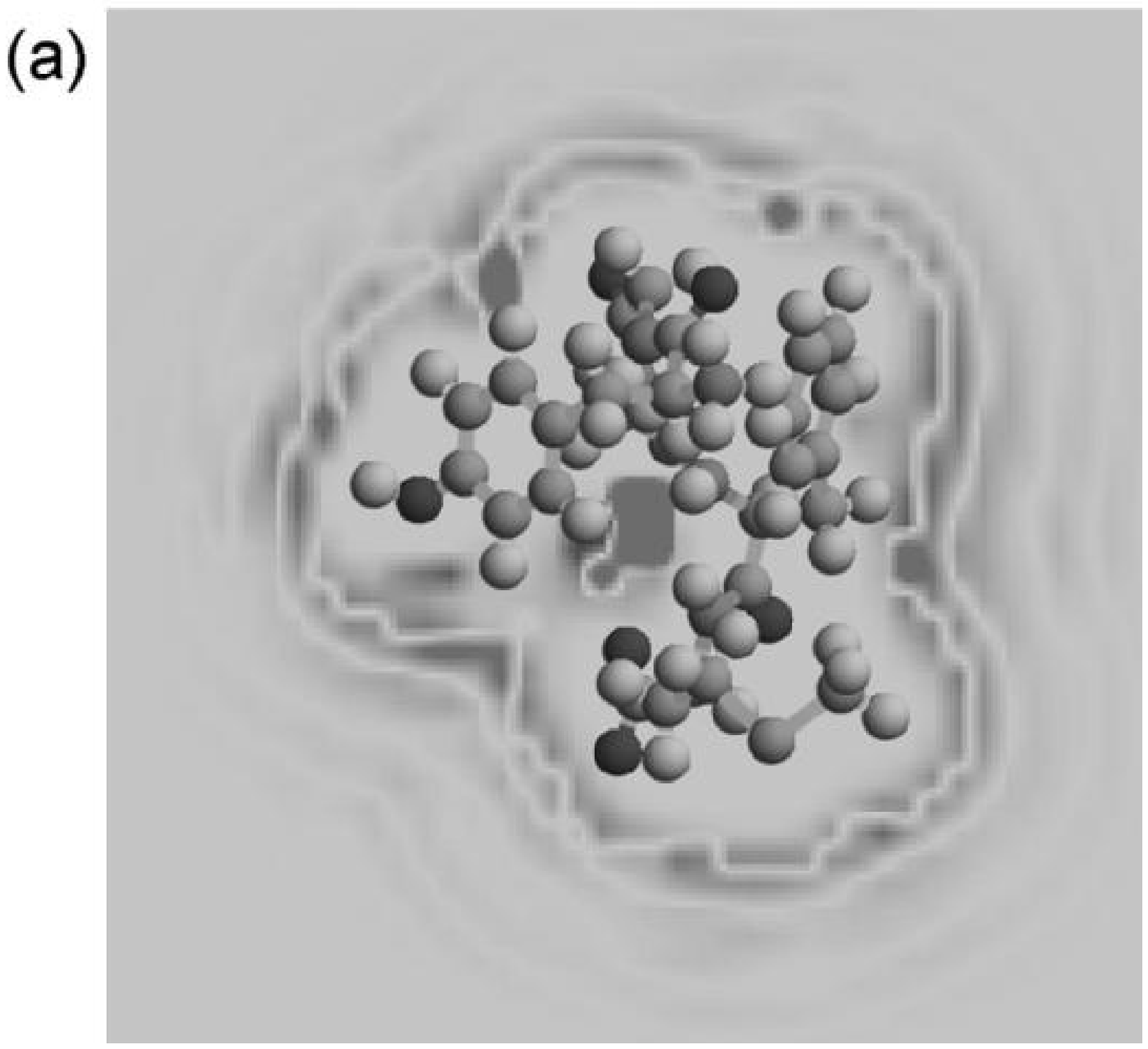}
\includegraphics[scale=0.25]{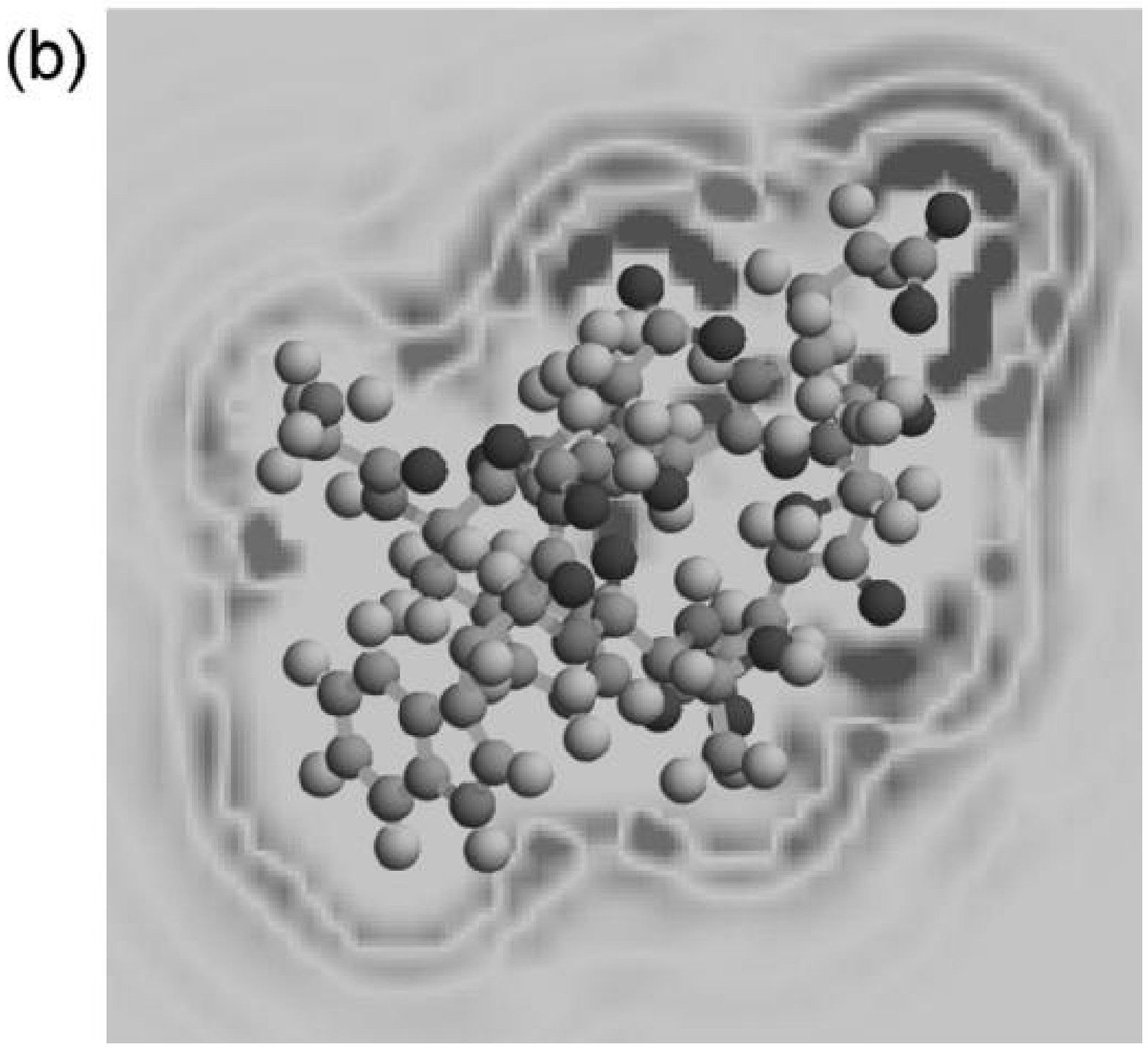}
}
\caption{Charge distribution of water around a solute molecule
obtained by 3D-RISM/FMO coupled simulations:
(a) methionine-enkephalin (1PLW, 75 atoms);
(b) chignolin (1UAO, 138 atoms).}
\label{fig:met-enkephalin}
\end{figure}

In the multi-physics simulations,
physical data are exchanged between separate program components,
where we must transform not only formats but also their semantics,
i.e., physical meanings of the data.
In order to assist such data-exchanges with semantic transformations,
we used a set of application program interfaces
called Mediator (mediator-API)\cite{mediator,NAREGI05},
which is included in the beta-version release
of the NAREGI grid-middleware.
Fig. \ref{fig:rism-fmo}(a) shows the total flow of this simulation,
where the partial charge distribution of the solute and solvent
molecules are exchanged each other through the mediator-API
(Fig. \ref{fig:rism-fmo}(b)).
In order to execute on the NAREGI grid,
the flow is incorporated in the NAREGI Workflow tool
(Fig. \ref{fig:rism-fmo}(c)).

In Fig. \ref{fig:met-enkephalin}, we show results of this
coupled calculation for methionine-enkephalin (75 atoms)
and chignolin (138 atoms) in aqueous solution,
where the partial charge distribution by water molecules are also shown
around these molecules.

\section{Development of OpenFMO Framework}

OpenFMO\cite{OpenFMO} is an open-licensed software platform
to construct FMO applications under high-performance
distributed computer environments.
The current status of this development is in the end of Phase II.
In Phase I, we introduced the OpenFMO framework
and predicted a peta-scale performance on a hypothetical
computer architecture\cite{HPCNano06}.
In Phase II, we have tried to implement the skeleton
by one-sided communications\cite{HPCAsia07}
under the PSI project\cite{PSI-Web}.
In Phase III, 
we are going to extend the platform to the multi-physics simulations
(see Fig. \ref{fig:OpenFMO}).

\subsection{Multi-physics Extension of OpenFMO and its Application}

\begin{figure}
\parbox{9cm}{
\begin{tabular}{|l|}
\hline
Phase I (2006)\\
\hspace{1cm}$\circ$ Fixing interfaces of MO library programs\\
\hspace{1cm}$\circ$ Skeleton with GAMESS-type algorithm\\
\hspace{1cm}$\circ$ Performance prediction for peta-computing\\
  Phase II (the first half of 2007)\\
\hspace{1cm}$\circ$ New skeleton for high-performance computing\\
\hspace{1cm}$\circ$ Performance analysis using MPI-2 functions\\
  Phase III (the latter half of 2007)\\
\hspace{1cm}$\circ$ Multi-physics extension\\
\hspace{1cm}$\circ$ Scientific applications\\
\hline
\end{tabular}
}\parbox{6.5cm}{
\includegraphics[scale=0.38]{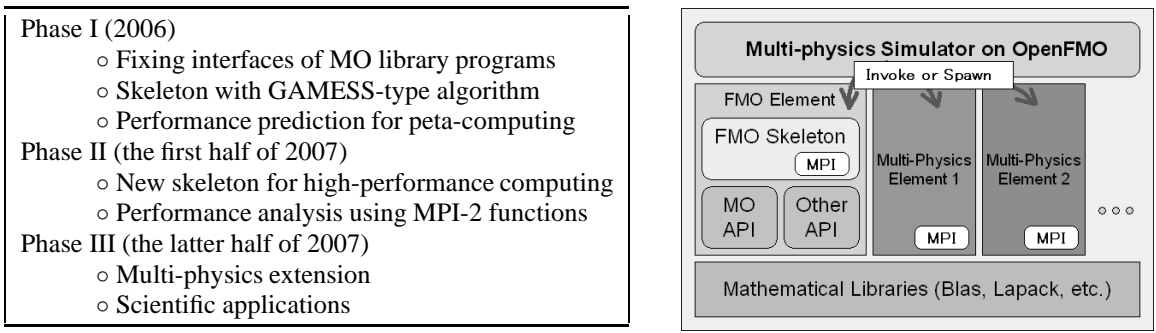}
}
\caption{Left: Development planning of OpenFMO.
Right: Stack structure of OpenFMO with multi-physics extension.
}
\label{fig:OpenFMO}
\end{figure}

The main purpose of the Phase II in the development schedule of OpenFMO
(Left of Fig. \ref{fig:OpenFMO}) was to correspond actual executions
on the next generation supercomputer with more than 10,000 CPUs,
where we have tried to reduce redundant memory consumption
on each computing nodes and improve parallel performance by the use of
one-sided communications.
The detailed results will be presented elsewhere\cite{HPCAsia07}.

The OpenFMO framework is extended to correspond to multi-physics
applications including scientific simulations (Fig. \ref{fig:OpenFMO}).
Since the current FMO skeleton is well configured
and has been proven effective in high-performance computing environments,
it is better that the other multi-physics components are developed separately.
Then, the key point is physical data representation and manipulation,
where sort of transparent accessibility
to internal data from outer components should be provided.
The semantic transformation of the data can also be executed
by the Mediator component depending on the needs.

By the use of the multi-physics extension,
one can construct the following coupled simulations:
QM/MM or QM/MD calculations\cite{QM/MM90,QM/MD03}
for docking simulations of proteins and enzymes in aqueous solution,
RISM/SCF simulations\cite{rismBook} for various protein molecules
with scientific theoretical studies\cite{TMO07}, etc.
One of the properties of the OpenFMO platform
is the lightweight and reconfigurable skeleton program,
which is useful for modifying applications
corresponding to fast-changing computational environments.

\section{Discussion}

Application sizes of recent molecular package programs are
more and more increasing by their self-contained structures
with many functions corresponding to complicated options
and various computer environments.
Such strategies in their development potentially run into a dead-end.
However, when we use multiple grid resources simultaneously,
it is inevitable to prepare a tailored scheduler\cite{Grid2007}
in each application.
The most important task as a scientist is to develop effective theories
and algorithms that can be implemented on every computer environments,
while actual implementation on a given environment
should be carefully done with the help of computer scientists.
Our development of the OpenFMO framework is one of those attempts
to implement the FMO algorithm on the future computers.


\vspace{0.6cm}
This work is partly supported by the Ministry of Education, Sports, Culture,
Science and Technology (MEXT) through the Science-grid NAREGI Program
under the Development and Application of Advanced High-performance
Supercomputer Project.



\bibliographystyle{aipproc}   

\bibliography{BibTeX/MyWorks,BibTeX/FMO}

\end{document}